\documentclass[sigconf]{acmart}
\usepackage{enumitem}
\usepackage{float}
\usepackage{subfig}
\AtBeginDocument{%
  \providecommand\BibTeX{{%
    \normalfont B\kern-0.5em{\scshape i\kern-0.25em b}\kern-0.8em\TeX}}}

\setcopyright{acmcopyright}
\copyrightyear{2020}
\acmYear{2020}
\acmDOI{10.1145/1122445.1122456}

\acmConference[San Diego '20]{San Diego '20: The26th SIGKDD Conference on Knowledge Discovery and Data Mining}{August 22--27, 2020}{San Diego, CA}
\acmBooktitle{San Diego '20: The 26th ACM SIGKDD Conference on Knowledge Discovery and Data Mining,
  August 22--27, 2020, San Diego, CA}
\acmPrice{15.00}
\acmISBN{978-1-4503-XXXX-X/20/08}

\newcommand{\blue}[1]{\textcolor{blue}{#1}}

\begin{document}

\title{HAN-ECG: An Interpretable Atrial Fibrillation Detection Model Using Hierarchical Attention Networks}
\author{Sajad Mousavi}
\affiliation{%
  \institution{Northern Arizona University}
  \city{Flagstaff}
  \state{Arizona}
  \postcode{86011}
}
\email{SajadMousavi@nau.edu}
\author{Fatemeh Afghah}
\affiliation{%
  \institution{Northern Arizona University}
  \streetaddress{8600 Datapoint Drive}
  \city{Flagstaff}
  \state{Arizona}
  \postcode{86011}}
\email{Fatemeh.Afghah@nau.edu}



\author{U. Rajendra Acharya}
\affiliation{%
 \institution{Ngee Ann Polytechnic}
 \country{Singapore}}
 \email{aru@np.edu.sg}




\renewcommand{\shortauthors}{Trovato and Tobin, et al.}

\begin{abstract}


Atrial fibrillation (AF) is one of the most prevalent cardiac arrhythmias that affects the lives of more than 3 million people in the U.S. and over 33 million people around the world, and is associated with a five-fold increased risk of stroke and mortality. Similar to other problems in healthcare domain, artificial intelligence (AI)-based algorithms have been used to reliably detect AF from patients' physiological signals. The cardiologist level performance in detecting this arrhythmia is often achieved by deep learning-based methods, however, they suffer from the lack of interpretability. In other words, these approaches are unable to explain the reasons behind their decisions. The lack of interpretability is a common challenge toward a wide application of machine learning (ML)-based approaches in the healthcare which limits the trust of clinicians in such methods. To address this challenge, we propose \textit{HAN-ECG}, an interpretable bidirectional-recurrent-neural-network-based approach for the AF detection task. The \textit{HAN-ECG} employs three attention mechanism levels to provide a multi-resolution analysis of the patterns in ECG leading to AF. The first level, wave level, computes the wave weights, the second level, heartbeat level, calculates the heartbeat weights, and third level, window (i.e., multiple heartbeats) level, produces the window weights in triggering a class of interest. The detected patterns by this hierarchical attention model facilitate the interpretation of the neural network decision process in identifying the patterns in the signal which contributed the most to the final prediction. Experimental results on two AF databases demonstrate that our proposed model performs significantly better than the existing algorithms. Visualization of these attention layers illustrates that our proposed model decides upon the important waves and heartbeats which are clinically meaningful in the detection task.

\end{abstract}




\maketitle

\section{Introduction}
Atrial fibrillation (AF) is a common cardiac arrhythmia that can lead to various heart-related complications such as stroke, heart failure and atrial thrombosis \cite{furberg1994prevalence}. Electrocardiography is a test which measures the electrical activity of the heart over a specific period of time. The test output is an Electrocardiogram (ECG) signal that is a plot of voltage against time. A common non-invasive diagnosis way for the AF detection is the process of the recorded electrocardiogram (ECG) signal visually by a cardiologist or medical
practitioner. However, this is a time-consuming process and subject to human error. 

Therefore, several computer-aided methods have been developed for automatic detection of atrial fibrillation and other heart arrhythmia. The existing ML-based methods include hand-crafted feature-based and automatic-extracted feature-based approaches in their solutions \cite{ghaffari2019atrial,xia2018detecting,acharya2017deep, asgari2015automatic,zhou2014automatic,lee2013atrial,lake2010accurate}. 
 Among them, the methods that extract features automatically have gained more attention because they could learn the ECG signal representations efficiently and achieve the state-of-the-art results. 
 
Machine learning approaches, especially deep learning techniques have achieved significant results in different applications ranging from  computer vision and reinforcement learning to unmanned aerial vehicle (UAVs) navigation \cite{xu2015show,mousavi2016learning, mousavi2017traffic, mousavi2016deep, mousavi2017applying, mousavi2018researching, shamsoshoara2019distributed, shamsoshoara2019solution, shamsoshoara2019autonomous} as well as healthcare domain \cite{rajpurkar2017cardiologist,mousavi2019sleepeegnet,xia2018detecting,zaeri2018feature}. Deep Learning models with the capability of automatic feature extracting provide significant performance in the AF detection task noting  their ability to detect complex patterns in the ECG signals \cite{yildirim2018arrhythmia,rajpurkar2017cardiologist}. Nevertheless, they work as black boxes that makes it hard to understand the reasons behind their decisions. Interpretability and transparency are key required factors in AI-based decision making  in the healthcare to enable and encourage the physicians who are held accountable for medical decisions to trust the recommendations of these algorithms. One way to make deep learning models interpretable is to incorporate an attention mechanism in the model that learns the relationship between the input data samples and the given task \cite{ma2017dipole}. 
To provide an interpretable method with high performance for automatic detection of the atrial fibrillation, in this study, we propose a deep learning model powered by hierarchical attention networks. The proposed method is composed of three parts in which each part contains a stacked bidirectional recurrent neural networks (BiRNN) followed by an attention model. The first part learns a wave level representation of the ECG signal, the second part learns a heartbeat level representation of the ECG signal and the third part learns a window-based (i.e., contains multiple heartbeats) level representation of the ECG signal. All learned representations at each level are interpretable and are able to show which parts of the input signal are the reasons to trigger an AF event. The performance of the proposed model was evaluated by two datasets of PhysioNet community, including the MIT-BIH AFIB database and the PhysioNet Computing in Cardiology Challenge 2017 dataset. The experiment results demonstrate that the proposed method significantly outperforms the state-of-the-art algorithms for the atrial fibrillation task. In addition, we show the interpretability of the learned representations in all levels through visualizations. The main contributions of this study are summarized as follows:
\begin{itemize}
    \item We propose an end-to-end hierarchical attention model that achieve the state-of-the-art performance with the capability of the interpretability.  
    \item The proposed  model provide a multi-level resolution interpretability (i.e., window by window (multiple heartbeats), heartbeat by heartbeat and wave by wave levels).
    \item We empirically demonstrate that the important parts of the ECG signal for the model in triggering the AF are clinically meaningful.
    \item The proposed approach can be used  to recognize new potential patterns leading to trigger life-threatening arrhythmias.
\end{itemize}

The hierarchical attention model was first proposed in \cite{yang2016hierarchical} in the content of document classification task, as a novel hierarchical attention architecture that matches the hierarchical nature of a document, meaning words make sentences and sentences make document. Since in the ECG analysis application, we deal with a similar notion of hierarchical input where the ECG signal includes multiple levels of resolution (waves, beats and windows)), the proposed hierarchical attention model can mirror the physicians’ decision-making process. For instance, in order to detect AF, they, first, look for some important windows (a sequence of continuous heartbeats), next, they look at the important heartbeats of the windows, and then focus on the heartbeat waves. 

The rest of this paper is organized as follows. Section \ref{sec:related-work} gives a review of the related work. Section \ref{sec:methodology} provides a detailed description of the proposed approach. Section \ref{sec:experiments} presents the experimental setup, the used databases to evaluate the proposed model, and compares the performance of the proposed algorithm to the existing algorithms following by the interpretability analysis. Finally, Section \ref{sec:conclude} concludes the paper.



\section{RELATED WORK}
\label{sec:related-work}
Life-threatening arrhythmia classification and prediction tasks are very important research problems in machine learning for healthcare area. Recent advances of deep learning algorithms have impacted on achieving great performance in the machine learning-oriented healthcare problems. Deep convolutional neural networks have been used to improve the performance of ECG heartbeat classification task \cite{kachuee2018ecg, acharya2017deep,jun2018ecg,yildirim2018arrhythmia}. Recurrent neural networks (RNNs) and sequence to sequence models were employed to perform automatic heartbeat annotations \cite{saadatnejad2019lstm, gao2019effective, mousavi2019inter}. 
Deep learning models have also been utilized to detect false arrhythmia alarms. \cite{lehman2018representation} applied a supervised senoising autoencoders (SDAE) to ECG signals to classify ventricular tachycardia alarms. \cite{mousavi2020single} used an attention-based
convolutional and recurrent neural networks to suppress false arrhythmia alarms in the ICU.

Atrial fibrillation (AF) is one of the most common type of arrhythmias in the patients with heart diseases and challenging arrhythmias to detect. \cite{fujita2019computer, ghaffari2019atrial,xia2018detecting} aimed to use  deep convolutional neural networks for the atrial fibrillation arrhythmia detection task and achieved good arrhythmia detection performance. \cite{shashikumar2018detection} applied an attention mechanism to detect the atrial fibrillation arrhythmia. Authors employed a deep recurrent neural network on 30-second ECG windows’ inputs, and also fed some time series covariates to the network. These covariates are hand-crafted features and include the standard deviation and sample entropy of the beat-to-beat interval time series. Although they have used an attention mechanism in the architecture of their model, their proposed method was not an interpretable detective model. The single-level attention applied on fixed-length 30s ECG windows, which contains several heartbeats only improves the detective performance. Our previous work named ECGNET \cite{mousavi2019ecgnet} is an interpretable atrial fibrillation detective model, which uses a deep visual attention mechanism to automatically extract features and focus on different parts of the heartbeats of the input ECG signal. The ECGNET has suggested an  interpretable AF detection with a single-level attention using the wavelet power spectrum as input, however, this study propose a hierarchical attention network having raw ECG signals as input. 

Unlike the aforementioned AF detective models, the proposed model provides a high resolution interpretable predictive model (i.e., window (i.e., multiple heartbeats) by window, beat by beat and wave by wave levels) using the hierarchical bidirectional recurrent neural networks and attention networks. The proposed model improve the detection performance and explain the reasons behind model decisions simultaneously.

\begin{figure*}[htb]
\centering
  \includegraphics[height=0.40\textheight,width=0.8\linewidth,keepaspectratio]{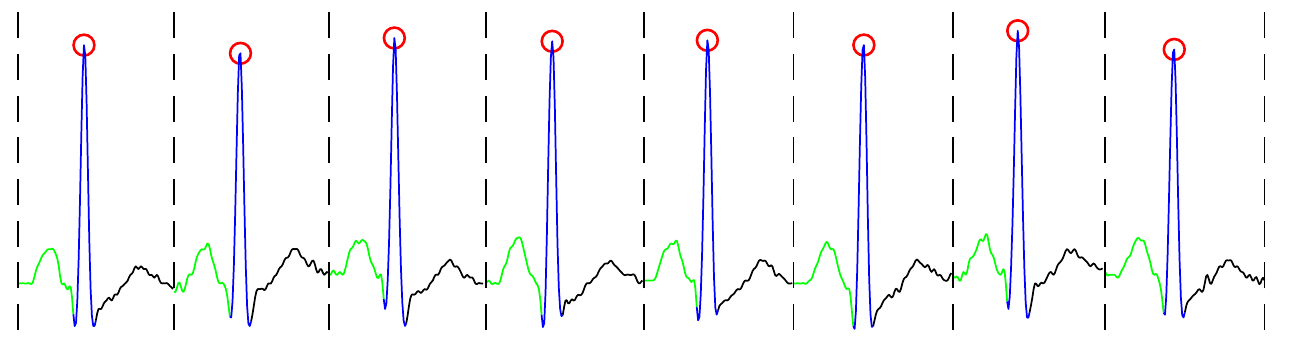}
  \caption{Illustration of an ECG signal; The red circles indicate R peaks; green, blue and black curves illustrate P, QRS and T-waves, respectively.} 
  \label{fig:ecg_wave}
\end{figure*}

\begin{figure*}[htb]
  \centering
  \includegraphics[width=0.8\linewidth,height=0.8\textheight]{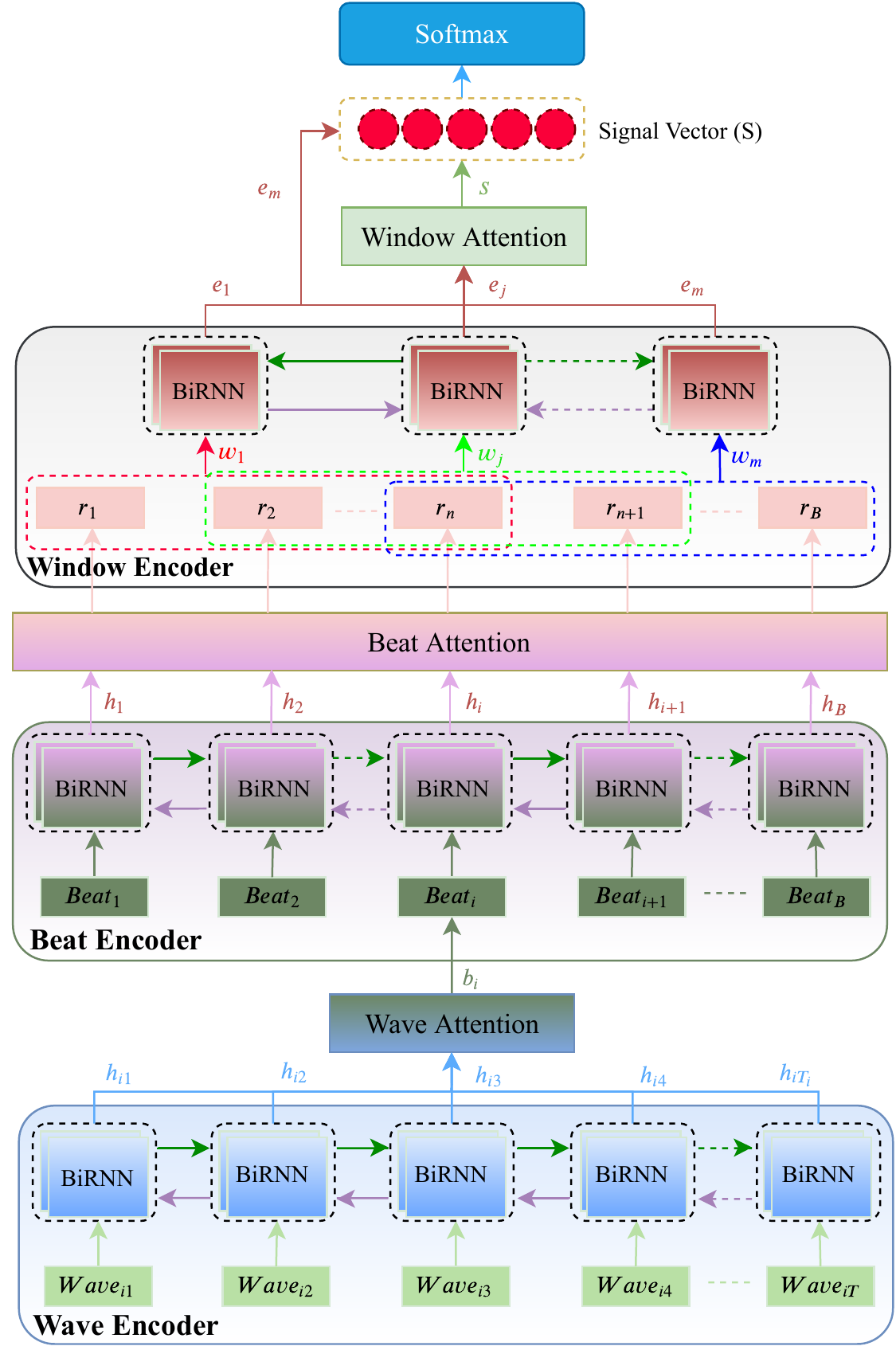}
  \caption{Architecture of hierarchical attention network (HAN) for Atrial Fibrillation Detection.}
  \label{fig:model-arch}
\end{figure*}
\begin{figure}[htb]
\centering
  \includegraphics[width=01.\linewidth,keepaspectratio]{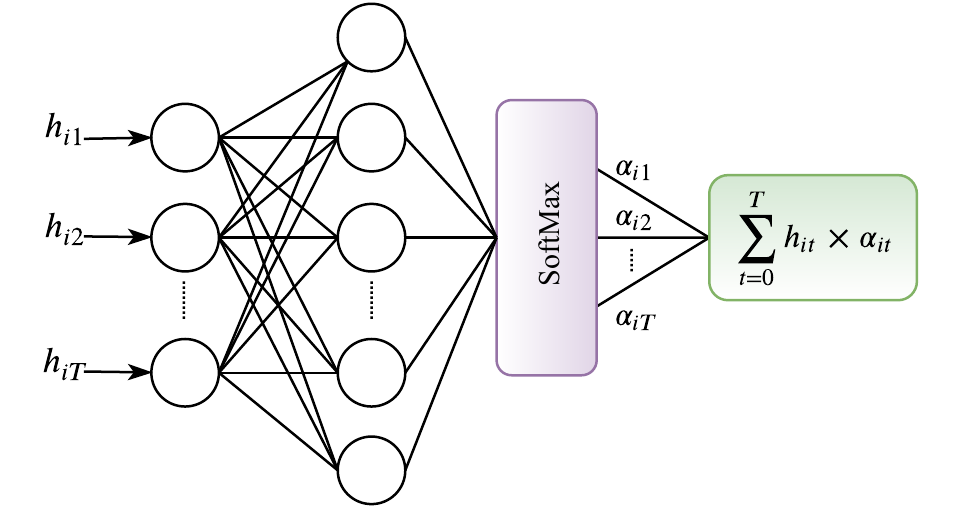}
  \caption{Schematic diagram of the attention mechanism.} 
  \label{fig:att-layer}
\end{figure}
\section{METHODOLOGY}
\label{sec:methodology}
In this section, we first describe the pre-processing steps needed to prepare the data to be fed into the proposed model. Then, we explain the main components of  the proposed method in detail.   

\subsection{Pre-processing}
\label{sec:per-processing}
The input of the proposed method is a sequence of ECG heartbeats in which each heartbeat contains a sequence of building waves (P-wave, QRS complex, T-wave, etc.). To prepare this structure of ECG signals, we perform a few pre-processing steps on them as follows:

\begin{enumerate} 
\item Removing the baseline wander and power-line interference noises in the ECG signal. To this end, the ECG signal was passed through a band-pass filter with a filter order of 10 and pass band of
0.5–50 Hz.
	\item Transforming the given ECG signal values to have a  zero mean and a unit standard deviation (i.e., standardization) 
\item Detecting the R-peaks of given ECG signal or detecting the QRS complexes using the Pan–Tompkins algorithm \cite{pan1985real}.
\item Dividing the continuous ECG signal into a sequence of heartbeats, and split the heartbeats into distinct units named waves. The waves in the ECG signal are extracted based on the extracted R-peaks and using adaptive searching windows, and a heartbeat is defined from the onset of the current P-wave to the offset of consecutive T-wave. Figure \ref{fig:ecg_wave} depicts a segmented ECG signal annotated with the R-peaks, P, QRS and T-waves.
\end{enumerate}
After doing the above pre-processing steps, each ECG signal becomes a sequence of $B$ heartbeats in which each heartbeat, $Beat_i$ contains $T_i$ waves, where $wave_{it}$ represents the $t^\text{th}$ ($t \in [1, T_i]$ wave in the $i^\text{th}$ ($i \in [1, B]$) heartbeat, $Beat_i$.

\subsection{Model}
The goal of the proposed method is to detect life-threatening atrial fibrillation arrhythmia in an explainable way. Figure \ref{fig:model-arch} presents the network architecture of the proposed method. A sequence of waves of an ECG signal is fed into a stacked bidirectional recurrent neural networks (BiRNN) followed by an attention model. The stacked BiRNNs are used to extract a vector representation for each input wave and the attention model is used to focus on those waves that are the best representatives of a heartbeat. Next, the vector representations of the waves are integrated to represent a heartbeat vector. Then, the heartbeat vectors of the previous step are introduced to other stacked BiRNNs followed by another attention model. Similarly, the attention model puts more emphasis on the important heartbeats and produces the heartbeat context vectors. After that, a sequence of windows in which each window contains multiple  heartbeat context vectors is computed and the same procedure is applied to the windows and a summarized vector that includes all information of the ECG signal is extracted. Finally, the summarized vector can be used for the atrial fibrillation detection task.  Overall, the model architecture is composed of three main parts: a wave encoder along with a wave attention, a beat encoder along with a beat attention, and a window encoder along with a window attention. In the following sections, we explain each part of the proposed model in detail.

\subsection*{Bidirectional Recurrent Neural Networks}
Bidirectional recurrent neural network (BiRNN) is more efficient than the RNN while the length of the sequence is very large \cite{schuster1997bidirectional}. The reason is that standard RNNs are unidirectional so they are restricted to only use the previous input state. However, the BiRNN can process data in both forward and backward directions. Therefore, the current state has access to previous and next input information simultaneously. In order to improve the detection performance, the BiRNNs were employed in the model for encoding the sequences of wave and beat vectors.

The BiRNN consists of forward and backward networks. The forward network takes in a sequence of waves/beats in a regular order, from $t=1$ to $t=T_i$, as input and computes forward hidden state, $\overrightarrow{h_t}$ and the backward network takes in wave/beat sequence in a reverse order, from $t=T_i$ to $t=1$, as input and calculates backward hidden state, $\overleftarrow{h_t}$. Then, the output of the BiRNN is considered as a weighted sum over the concatenation of the forward hidden state, $\overrightarrow{h_t}$ and the backward one, $\overleftarrow{h_t}$. The BiRNN can be defined mathematically as follows:
\begin{align}
  &\begin{aligned} 
    \overrightarrow{h_t}= \tanh(\overrightarrow{W}x_t+\overrightarrow{V}{\overrightarrow{h}}_{t-1}+\overrightarrow{b})
  \end{aligned}\\
  &\begin{aligned}
   \overleftarrow{h_t}= \tanh(\overleftarrow{W}x_t+\overleftarrow{V}{\overleftarrow{h}}_{t+1}+\overleftarrow{b})
  \end{aligned} \\
    &\begin{aligned}
  y_t = (U[ \overrightarrow{h_t}; \overleftarrow{h_t}] + b_y),
  \end{aligned}
\end{align}
where ($\overrightarrow{h_t}$, $\overrightarrow{b}$) are the hidden state and the bias of the froward network, and ($\overleftarrow{h_t}$, $\overleftarrow{b}$) are the hidden state and the bias of the backward one. In addition, $x_t$ and $y_t$ are the input and the output of the BiRNN, respectively. 
\subsection*{Wave Encoder and Wave Attention}
A sequence of waves, $wave_{it}$ $t \in [1, T_i]$, for $i^\text{th}$ heartbeat, is fed into a bidirectional recurrent neural network (BiRNN) to encode the wave sequence. The forward network of the BiRNN gets the heartbeat, $i$ in a normal time order of waves from $wave_{i1}$  to $wave_{iT}$ and the backward network gets the heartbeat, $i$ in a reverse time order of waves from $wave_{iT}$  to $wave_{i1}$. Then, the BiRNN outputs, $h_{it}$ representing a low dimensional latent vector representation of the heartbeat, $i$. 

Similar to the words of a sentence in which necessarily all words are not important to give the meaning of the sentence \cite{yang2016hierarchical}, herein all waves of a heartbeat do not have the same weights in representing the heartbeat. Therefore, an attention mechanism is able to extract the relevant waves of the heartbeat that contribute more to the meaning of the heartbeat. The attention mechanism is a shallow neural network that takes the BiRNN output, $h_{it}$ as input and computes a probability vector, $\alpha_{it}$ corresponding to the importance of each wave vector. Then, it calculates a wave context vector, $b_i$ which is a weighted sum over $h_{it}$ with the weight vector $\alpha_{it}$ (as shown in Figure \ref{fig:att-layer}). Indeed,
\begin{align}
  &\begin{aligned} 
    \alpha_{it}= softmax(V_w \tanh(W_wh_{it}+b_w))
  \end{aligned}\\
  &\begin{aligned} 
    b_i=\sum_{t}{\alpha_{it}h_{it}},
  \end{aligned}
\end{align}
where $(W_w,b_w,V_w)$ are the parameters to be learned and $softmax(.)$ is a function that squeezes its input, which is a vector of real numbers, in values between 0 and 1.

\subsection*{Beat Encoder and Beat Attention}
Similar to the wave encoder part, the BiRNN of the beat encoder part takes a sequence of wave context vectors, $b_i \; (i \in [1, B])$ as input and produces vectors, $h_i \; (i \in [1, B])$ which are latent representations of the input heartbeats. To put emphasise on the more important heartbeats in triggering the arrhythmia, another attention mechanism is used on the heartbeat level. Therefore, 

\begin{align}
  &\begin{aligned} 
    \alpha_{i}= softmax(V_b \tanh(W_bh_{i}+b_b))
  \end{aligned}\\
  &\begin{aligned} 
    r=\alpha_{i} \circ h_{i},
  \end{aligned}
\end{align}
where $(W_b,b_b,V_b)$ are the parameters to be learned, $\alpha_{i}$ is the attention weight vector of the heartbeats, and $r \; (r_1, r_2, \ldots, r_B)$ is the heartbeat context vectors which is an element-wise product of the hidden states, $h_{i}$ and the importance of each heartbeat, $\alpha_{i}$.

\subsection*{{Window Encoder and Window Attention}}
In addition to the wave and heartbeat level encoding modules, we also consider a window level encoding module in which a window contains multiple heartbeats. The heartbeat context vectors, $(r_1, r_2, \ldots, r_B)$ are converted to a sequence of windows, $w_{j} \; (j = 1,2, \ldots, m)$ by sliding a predefined-fixed-length window  with a predefined-fixed hop size in the heartbeats over the heartbeat context vectors (as shown in Figure \ref{fig:model-arch}). For example, if we consider a window with $n=3$ heartbeats and the hop size be 1, the extracted sequence of windows becomes $(w_1= [r_1;r_2;r_3], \; w_2 = [r_2;r_3;r_4], \; \ldots \;, \; w_m= [r_{B-2};r_{B-1};r_B])$ where $;$ is a simple concatenation operation.

Analogous to the previous steps, a BiRNN is used to encode the windows, $w_{j} \; (j = 1,2, \ldots, m)$ and again  an attention mechanism is employed to measure the
importance of the windows. Specifically,  
\begin{align}
  &\begin{aligned} 
    \gamma_{j}= softmax(V_v \tanh(W_ve_{j}+b_v))
  \end{aligned}\\
  &\begin{aligned} 
    s=\sum_{j}{\gamma_{j}e_{j}},
  \end{aligned}
\end{align}
where $(W_v,b_v,V_v)$ are the parameters to be learned, $\gamma_{j}$ is the attention weight vector of the windows, $e_{j} \; (j = 1,2, \ldots, m) $ are the hidden states of the BiRNN, and $s$ is the window context vector that encompasses the whole information of the windows, containing multiple heartbeats, of the input ECG signal.

\subsection*{Detection}
We concatenate the window context vector, $s$ and the last hidden state, $e_m$, to obtain a combined information of both vectors and then feed it into a shallow network followed by a softmax layer to produce a probability vector, $p$ in which each element determines the probability of the input signal belonging to each class of interest (AF or non-AF). Specifically,
\begin{align}
  &\begin{aligned} 
    S = \tanh(W_c[s;e_m])
  \end{aligned}\\
  &\begin{aligned} 
    p=softmax(WsS+b_s),
  \end{aligned}
\end{align}
where $(W_c,Ws,b_s)$ are the parameters to be learned.

Finally, we use a cross entropy loss to calculate the training loss as follows:
\begin{equation}
L=-y \cdot \log p,
\end{equation}
where $(\cdot)$ is the vector dot product operator and $y$ is  the ground truth vector.

\subsection*{Interpretation}
Typically artificial intelligence (AI)-based algorithms that both give good performance and are interpretable, are preferable to apply to real medical practice. Therefore, having machine learning algorithms that explain the reasons behind their decisions are very important in medical applications. The proposed method has three levels of the attention mechanism, the first level (i.e., the wave level) produces the wave weights, $\alpha_{it} \; (t=1,2,...,T_i$)  representing the importance of the waves in a heartbeat, the second level (i.e., the heartbeat level) computes the heartbeat weights, $\alpha_i \; (i=1,2,...,B$) showing the amount of the influence of each heartbeat on the occurrence of an arrhythmia, and third level (i.e., the window level) produces the window weights, $\gamma_j \; (j=1,2,...,m$) demonstrating the importance of the combinations of the heartbeats. In Section \ref{sec:visu}, we provide visualized examples of some ECG signals with the AF and non-AF arrhythmias  where the focused portions of the signals determined by the proposed attention mechanism are highlighted. 

\section{EXPERIMENTS}
\label{sec:experiments}
In this section, we describe the two atrial fibrillation datasets used for the quantitative and qualitative analyses of the proposed method. Then, we compare its performance against the existing algorithms for the atrial fibrillation detection task and show how explainable the proposed model is in detecting the atrial fibrillation arrhythmia.
\subsection{Data Description}
To evaluate the proposed method, we used two datasets including the MIT-BIH AFIB database \cite{PhysioNetmitafdb} and  the PhysioNet Computing in Cardiology Challenge 2017 dataset \cite{PhysioNetafdb17}. 

\textbf{MIT-BIH AFIB Dataset:} This dataset contains 23 long-term ECG recordings of human subjects with mostly atrial fibrillation arrhythmia. Each patient of the MIT-BIH AFIB includes two   10-hours long ECG recordings (ECG1 and ECG2). The ECG signals are sampled at 250 Hz with 12-bit resolution over a range of $\pm10$ millivolts. In this study, we divided each ECG signal into 5-s segments and labeled each based on a threshold parameter, $p$. To perform the segment labeling, we followed the approach reported in \cite{xia2018detecting,asgari2015automatic}. A 5-s segment is labeled as AF if the percentage of annotated AF beats of the segment is greater than or equal to $p$, otherwise it is determined as a non-AF arrhythmia. We chose $p=50\%$ to be consistent with the previous research work. In our experiments, we used the ECG1 recordings and extracted a total of 167,422 5-s data segments in which the the number of AF segments was 66, 939 and the number of non-AF segments was 100, 483. As it is clear, the data segments are imbalanced. To deal with this problem and be able to compare our proposed method to the other existing algorithms, we randomly drew the same number of samples for both AF and non-AF classes (considered 66, 939 samples for each class). However, we tested the proposed method on the original imbalanced dataset as well.   

\textbf{PhysioNet Challenge AFIB Dataset:} The goal of the challenge is to build the models to classify a single short ECG lead recording (30-60s in length) to normal sinus rhythm, atrial fibrillation (AF), an alternative rhythm, or too noisy classes. The training set includes 8,528 single lead ECG recordings and the test set contains 3,658 ECG recordings. The test set has not been publicly available yet, therefore we use the training set for both test and training phases. The ECG recordings were recorded by AliveCor devices,  sampled as 300 Hz and filtered by a band pass filter. In this study, we considered only two classes including the normal sinus rhythm (N) and atrial fibrillation (AF), and discarded the remaining groups.

\subsection{Experimental Setup}
The proposed approach is based on the hierarchical attention networks and has employed three levels of attention. To show the performance of this proposed model, in our experiments, we  consider the model without the attention mechanism (denoted as RNN containing just the BiRNNs), one\blue{-} (denoted as HAN-ECG\textsubscript{1}), two\blue{-} (denoted as HAN-ECG\textsubscript{2}) and three- levels (denoted as HAN-ECG\textsubscript{3}) of the attention mechanism. 

We applied a 10-fold cross-validation approach to evaluate the model. Indeed, we split the dataset into 10 folds. At each round of the cross validation, 9 folds were used for training the model and the  remaining fold (1 fold) was used for evaluating the model. At the end, we combined all the evaluation results. 

The models were trained with a maximum of 25 epochs and mini-batches of size 64. The Adam optimizer was used to minimize the loss, $L$ with a learning rate $\alpha=0.001$. We also used a $L_2$ regularization with a coefficient $\beta = 1e-5$ and a drop-out technique with a probability of dropping of 0.5 to reduce the effect of the overfitting problem during the training. The number of layers for the BiRNNs ware set to 2. The window and the hop sizes for the last attention layer were set to (2,2) and (5,5) for the MIT-BIH AFIB and AFDB17 databases, respectively.
We utilized Python programming language and Google Tensorflow deep learning library to implement the proposed model. We ran the 10-fold cross-validation on a machine with 8 CPUs (Intel(R) Xeon(R) CPU @ 3.60 GHz), 32 GB memory and Ubuntu 18.04. In all experiments, the best performance were reported.

\subsection{Results}
\subsection*{\normalsize{Quantitative analysis}}
Table \ref{tab:compare} shows the performance of the proposed method with different numbers of employed attention mechanisms against the state-of-the-art algorithms on the MIT-BIH AFIB database with the ECG segment of size 5-s. It can be seen from the table that the proposed method with one, two and three hierarchies achieved significantly better performance against other methods listed in the table. In Table \ref{tab:compare}, we can observe that the accuracy of the proposed method with two levels of attention is slightly higher than the one with three levels. The reason might be that the ECG segment of size 5-s (as input) has approximately 6 heartbeats in which almost all heartbeats contains the AF arrhythmia. Therefore, the heartbeats windowing at the level three makes no significant improvement in the model performance.  

\begin{figure*}[ht]%

    \centering
    \subfloat[RNN]{\includegraphics[width=6cm]{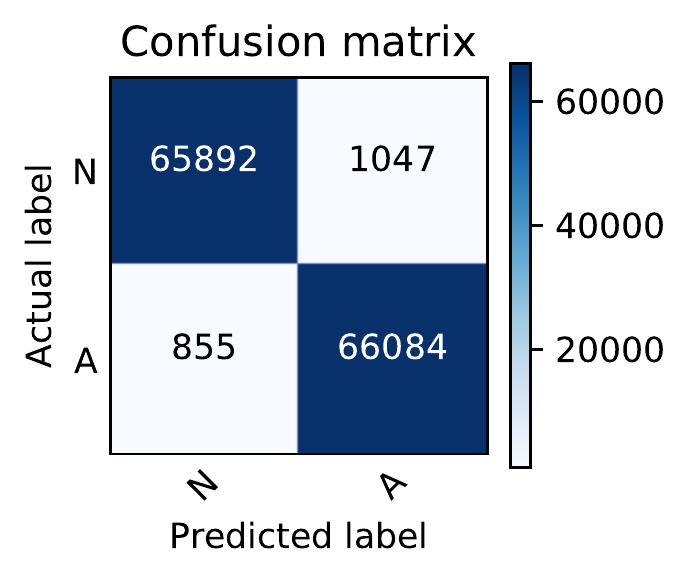} 
    }%
    \quad
     \subfloat[HAN-ECG\textsubscript{1}]{\includegraphics[width=6cm]{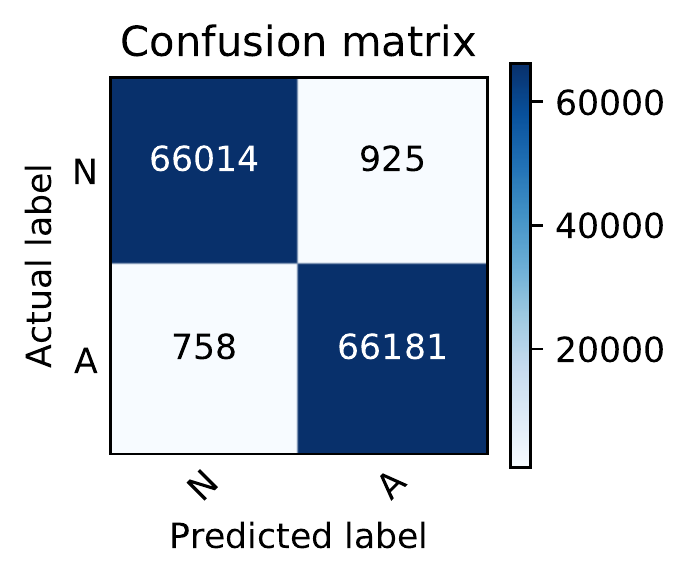}  
      }%
         \quad
     \subfloat[HAN-ECG\textsubscript{2}]{\includegraphics[width=6cm]{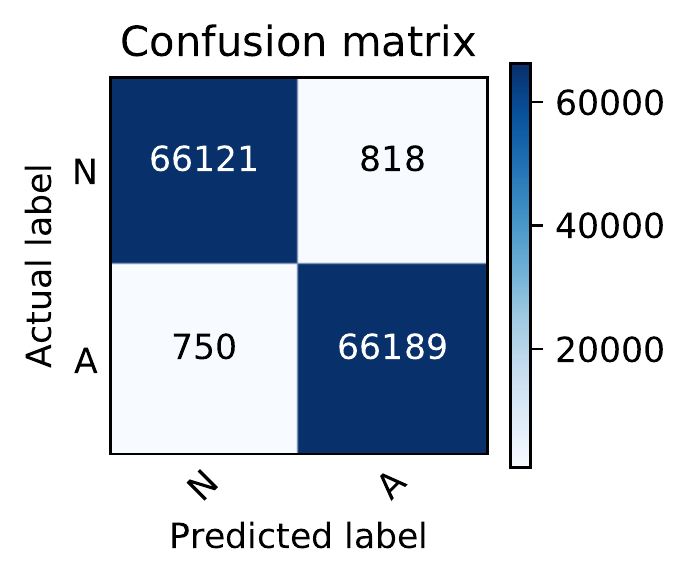}  
      }%
          \quad
     \subfloat[HAN-ECG\textsubscript{3}]{\includegraphics[width=6cm]{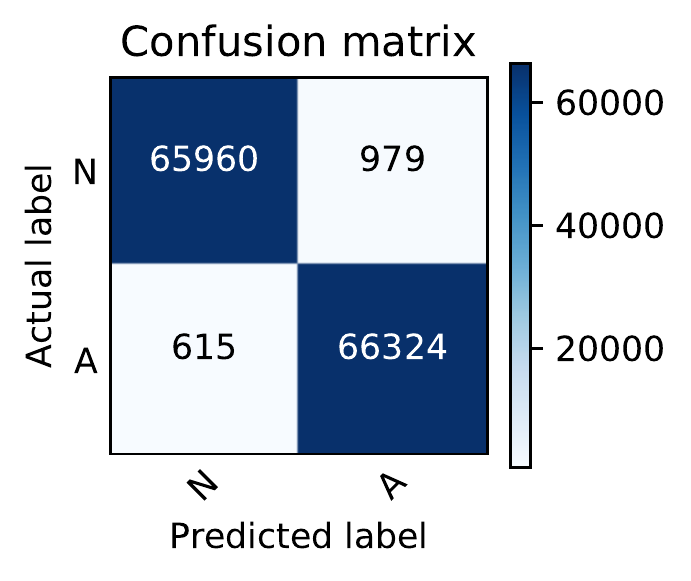}  
      }%
    \caption{Confusion matrices achieved by all the proposed method variants on the MIT-BIH AFIB database.}%
    \label{fig:cms}%

\end{figure*}

The row number 5 (i.e., the method named HAN-ECG\textsubscript{2f})  in Table \ref{tab:compare} shows the evaluation results of the proposed method while the input ECG signals are split into fixed size portions (here 180 samples for each portion as a heartbeat) and the portions are divided into fixed-size parts (here 6 parts for each portion and each part is considered as a distinc T-wave). From Table \ref{tab:compare}, it is clear that the proposed method using the fixed size heartbeats and the waves as input results in the lowest performance among all the proposed method variants. Hence, we can conclude that the pre-processing step in our methodology, as shown in Section \ref{sec:per-processing}, is necessary to obtain better performance. It can also be seen that the RNN method can preform as good as the algorithm provided by Xia et al \cite{xia2018detecting} which is a deep convolutional neural network with the stationary wavelet transform (SWT) coefficient time series as input. In addition, Figure \ref{fig:cms} illustrates the confusion matrices' plots to describe a summary of how well the proposed model is performing given all folds. 

\begin{table} [htb] 
\caption{Comparison of performance of the proposed model against other algorithms on the MIT-BIH AFIB database with the ECG segment of size 5-s.}
 \centering{
\label{tab:compare}
	\resizebox{1.0\linewidth}{!}{  
\begin{tabular}{ccccccc}
\toprule
\textbf{} & \textbf{} & \multicolumn{4}{c}{\textbf{Best Performance}} \\
\cmidrule(lr){3-6}
\textbf{Method} &  \textbf{Database} &  {$Sensitivity$} & {$Specificity$} & {$Accuracy$} & {$AUC$}  \\
\midrule
HAN-ECG\textsubscript{3} & AFDB  &$\textbf{99.08}$ &${98.54}$ &${98.81}$ &$\textbf{99.86}$ \\
HAN-ECG\textsubscript{2} & AFDB  &${98.88}$ &$\textbf{98.78}$ &$\textbf{98.83}$ &${99.85}$ \\
HAN-ECG\textsubscript{1} & AFDB  &${98.87}$ &${98.62}$ &${98.74}$ &${99.84}$ \\
RNN & AFDB  &${98.72}$ &${98.44}$ &${98.58}$ &${99.80}$ \\
HAN-ECG\textsubscript{2f} & AFDB  &${98.68}$ &${98.36}$ &${98.52}$ &${99.79}$ \\
Xia et al. (2018) \cite{xia2018detecting} &AFDB  &$98.79$ &$97.87$ &$98.63$ &$-$ \\
Asgari et al. (2015) \cite{asgari2015automatic} &AFDB  &$97.00$ &$97.10$ &$-$ &$-$ \\
Lee, et al. (2013) \cite{lee2013atrial} &AFDB  &$98.20$ &$97.70$ &$-$&$-$  \\
Jiang et al. (2012) \cite{jiang2012high} &AFDB  &$98.20$ &$97.50$ &$-$ &$-$ \\
Huang et al. (2011) \cite{huang2011novel} &AFDB  &$96.10$ &$98.10$ &$-$&$-$  \\
Babaeizadeh, et al. (2009) \cite{babaeizadeh2009improvements} &AFDB  &$92.00$ &$95.50$ &$-$ &$-$ \\
Dash et al. (2009) \cite{dash2009automatic} &AFDB  &$94.40$ &$95.10$ &$-$ &$-$ \\
Tateno et al. (2001) \cite{tateno2001automatic} &AFDB  &$94.40$ &$97.20$ &$-$ &$-$ \\
 \bottomrule  
\end{tabular} }
}
\end{table}

The reported results by our proposed method and other listed algorithms in Table \ref{tab:compare} are based on balancing the dataset before training the models, in which the same number of non-AF data samples as the AF data samples are selected randomly. In addition, the selection of the 5-s data segments is from all combined data segments extracted from all individuals. Therefore, the training and evaluation sets can include data segments from the same subjects which is a data leakage problem. To have a more realistic evaluation mechanism, we considered another scenario in which the test and training data segments came from different individuals, and left the dataset imbalanced. Table \ref{tab:compare2} presents the performance of the proposed AF detectors with the new evaluation scenario on the MIT-BIH AFIB database. Since we could not find any research paper that followed the aforementioned scenario, we just reported our results without any comparison in Table \ref{tab:compare2}. From Table \ref{tab:compare2}, we can again note that the models with more attention layers yield higher accuracy and better performance.


\begin{table} [h] 
\caption{Performance of the proposed model for the AF classification task on the MIT-BIH AFIB database while the database is not balanced and the data segments for the training and test phases come from different ECG recordings.}
 \centering{
\label{tab:compare2}
	\resizebox{1.0\linewidth}{!}{  
\begin{tabular}{ccccccc}
\toprule
\textbf{} & \textbf{} & \multicolumn{4}{c}{\textbf{Best Performance}} \\
\cmidrule(lr){3-6}
\textbf{Method} &  \textbf{Database} &  {$Sensitivity$} & {$Specificity$} & {$Accuracy$} & {$AUC$}  \\
\midrule
HAN-ECG\textsubscript{3} & AFDB  &${90.53}$ &${79.54}$ &${82.41}$ &${89.46}$ \\
HAN-ECG\textsubscript{2} & AFDB  &${89.86}$ &${77.49}$ &${81.58}$ &${88.65}$ \\
HAN-ECG\textsubscript{1} & AFDB  &${89.47}$ &${75.15}$ &${79.96}$ &${85.94}$ \\
RNN & AFDB  &${89.20}$ &${74.38}$ &${79.55}$ &${85.88}$ \\
 \bottomrule  
\end{tabular} }
}
\end{table}

Table \ref{tab:afcompare} shows the experimental results on the the PhysioNet Computing in Cardiology Challenge 2017 dataset. The overall performance of the proposed models with more than one attention layer (i.e., HAN-ECG\textsubscript{2} and HAN-ECG\textsubscript{2}) is better than other methods, demonstrating the hierarchical attention networks works better for the AF detection task. Since, in this experiment, we considered a two-class problem (AF and normal classes), there was not any work in the literature to report a comparison. 
\begin{table} [h] 
\caption{Performance of the proposed model for the AF classification task on the PhysioNet Computing in Cardiology Challenge 2017 dataset (AFDB17). }
 \centering{
\label{tab:afcompare}
	\resizebox{1.0\linewidth}{!}{  
\begin{tabular}{ccccccc}
\toprule
\textbf{} & \textbf{} & \multicolumn{4}{c}{\textbf{Best Performance}} \\
\cmidrule(lr){3-6}
\textbf{Method} &  \textbf{Database} &  {$Sensitivity$} & {$Specificity$} & {$Accuracy$} & {$AUC$}  \\
\midrule
HAN-ECG\textsubscript{3} & AFDB17  &${86.02}$ &${98.62}$ &${96.98}$ &${98.46}$ \\
HAN-ECG\textsubscript{2} & AFDB17  &${86.15}$ &${98.50}$ &${96.90}$ &${98.41}$ \\
HAN-ECG\textsubscript{1} & AFDB17  &${84.30}$ &${98.48}$ &${96.64}$ &${98.44}$ \\
RNN & AFDB17  &${80.52}$ &${97.40}$ &${95.18}$ &${97.18}$ \\
 \bottomrule  
\end{tabular} }
}
\end{table}

\subsection*{\normalsize{Qualitative  analysis}}
\label{sec:visu}
\begin{figure*}[htb]
  \centering
  \includegraphics[width=1.0\linewidth,]{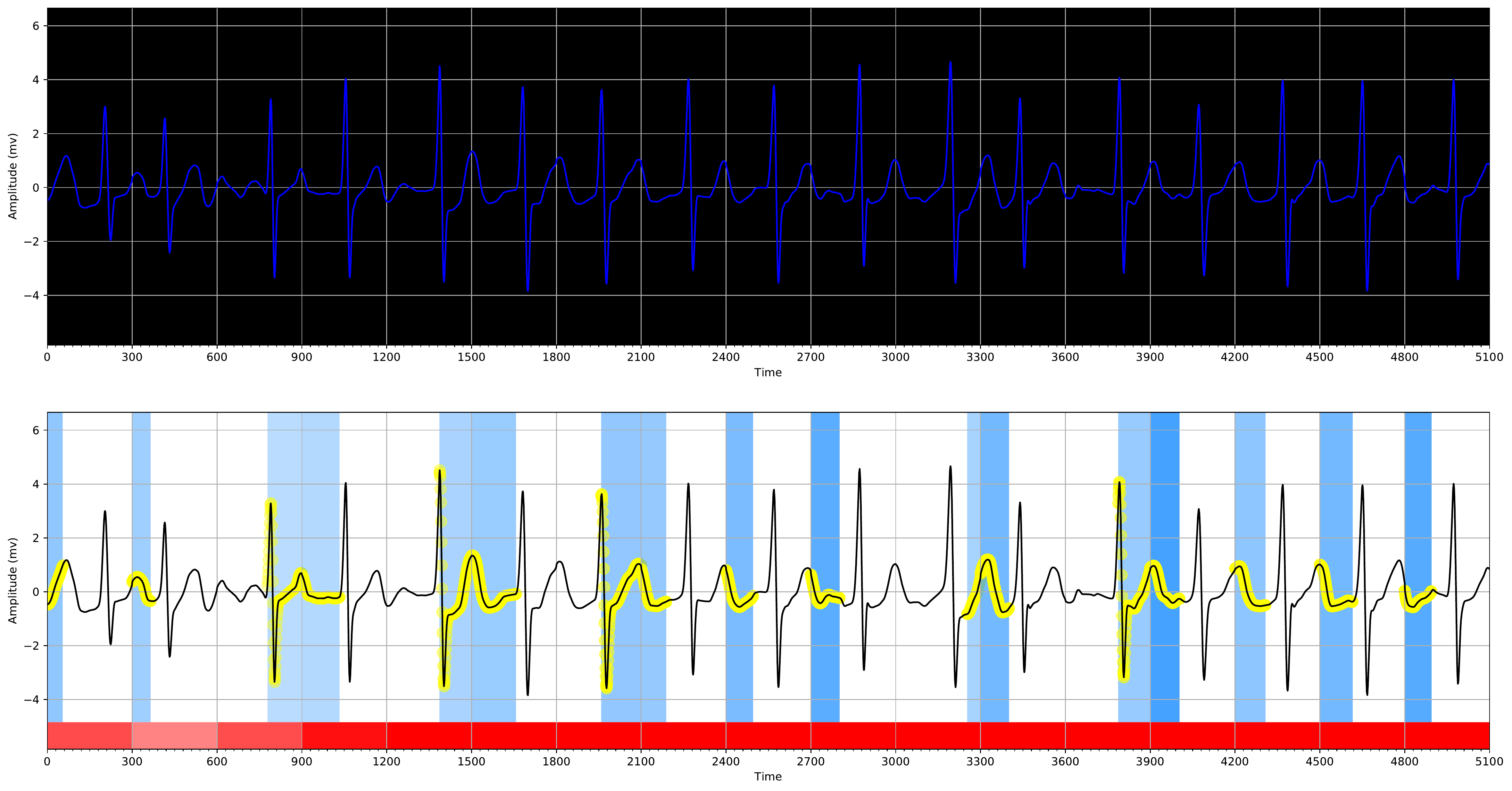}
  \caption{Hierarchical attention visualisation of a subject with the AF arrhythmia from the PhysioNet Computing in Cardiology Challenge 2017 dataset.}
  \label{fig:attAA1}
\end{figure*}

\begin{figure*}[ht]%

    \centering
        \subfloat[Subject without the AF arrhythmia]{\includegraphics[width=.49\linewidth]{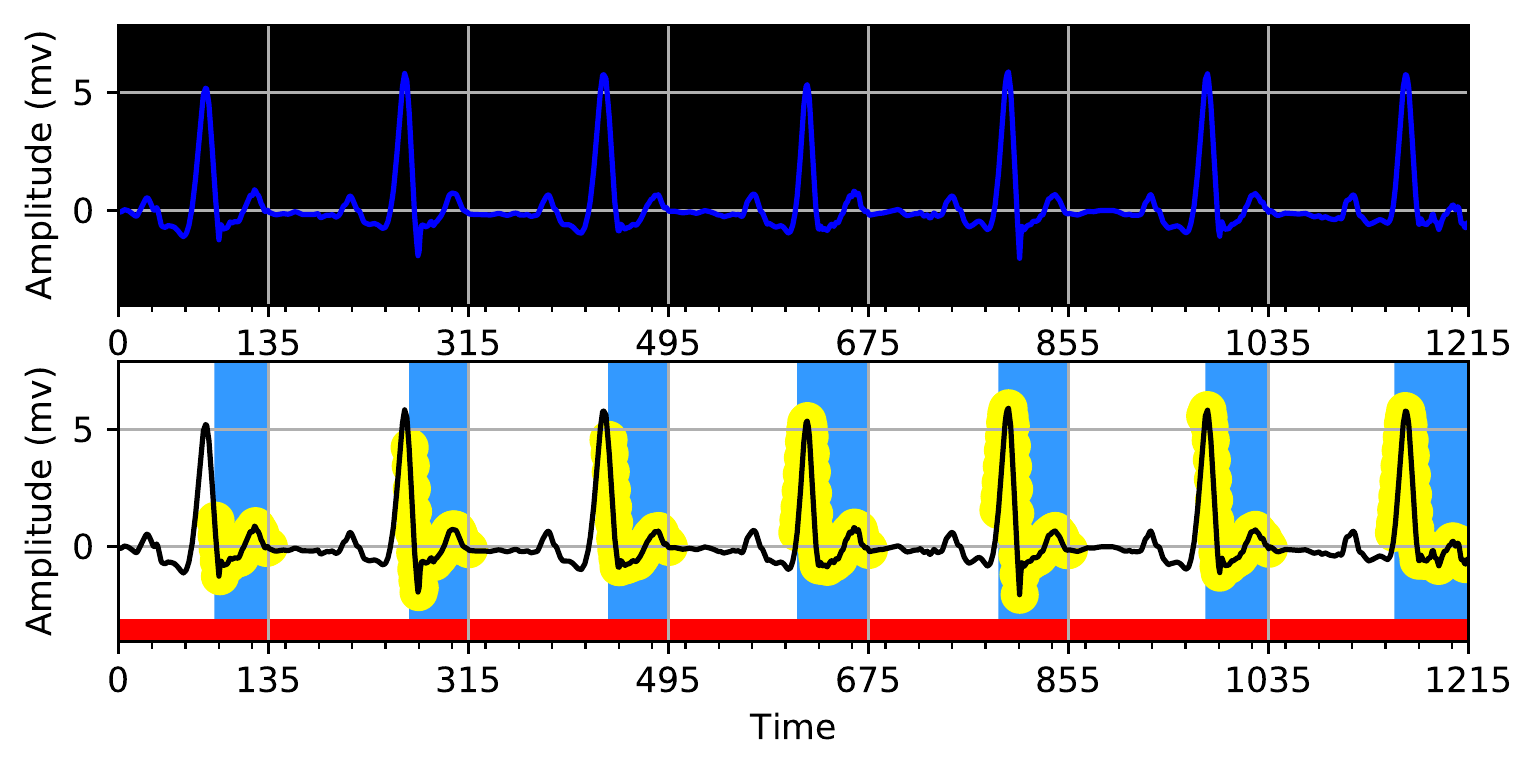} 
      \label{fig:att_mit_af}
    }%
      \subfloat[Subject with the AF arrhythmia]{\includegraphics[width=.49\linewidth]{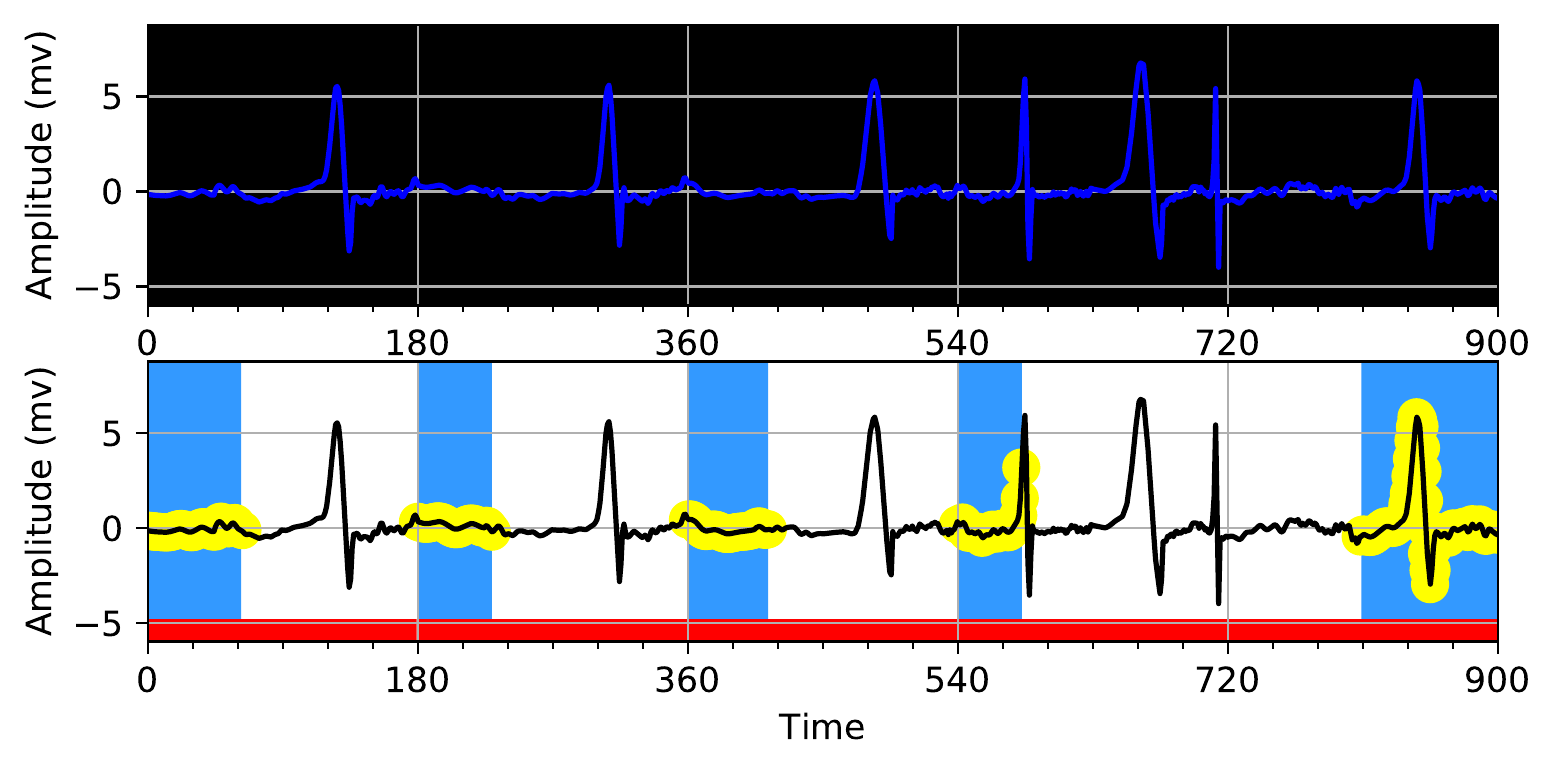}  
 \label{fig:att_mit_naf}
      }%

    \caption{ Hierarchical attention visualisation of two subjects from the MIT-BIH AFIB database.}
    \label{fig:att_mit}%

\end{figure*}

Understanding the cause of the model decision is very important in healthcare applications. In order to validate that the decisions made by our model are interpretable, we demonstrate through visualising the hierarchical attention layers that the proposed method is considering clinically important heartbeats and waves in detecting the AF arrhythmia. Figures \ref{fig:attAA1} and \ref{fig:att_mit} illustrate a few ECG signals containing the AF and non-AF categories. The top plots of the figures show the original ECG signals and the bottom plots present the informative heartbeats and waves in the detection of a class of interest (AF and non-AF). In the figures, the red segment denote the heartbeat weights and darker ones show more important heartbeats on the network's decision, and the blue strips and the yellow circles denote the locations of the important waves of the heartbeats in which the darker blue ones show more influence on the detection process. 

There are two essential visual features in the patient ECG signals of that the practitioners use to identify the atrial fibrillation: (i) the absence of P-waves that occasionally are replaced by a series of small waves called fibrillation waves, and (ii) the irregular R-R intervals in which the heartbeat intervals are not rhythmic. Figure \ref{fig:attAA1} visualizes the important regions of the ECG signal while it contains AF arrhythmia. From the figure, we can see the importance of the heartbeats (i.e., through the intensity of the red segments), and that the proposed method pays attention to the irregularity of R-R intervals and emphasizes on the absence of P-waves which are the clinical features in recognizing the atrial fibrillation.

In addition, the proposed hierarchical attention mechanism considers the normal heartbeat rhythms for the detection of the non-AF class as shown in Figure \ref{fig:att_mit_af}. In order to label an ECG signal as the AF, from Figure \ref{fig:att_mit_naf} we can again observe that  the model is interested in the parts of the ECG signal in which the P-waves are absent (replaced with a series of low-amplitude oscillations). Since all the heartbeats of the 5-s data segments in Figure \ref{fig:att_mit_af} and \ref{fig:att_mit_naf} are either the normal heartbeats or the atrial fibrillation heartbeats, the importance of all the heartbeats approximately is the same (i.e., the same intensity for the red segments). Generally, in all aforementioned figures, our proposed model considers the clinically meaningful waves and their corresponding heartbeats in its decision-making process.

\section{Conclusions}
In this study, we proposed a hierarchical attention mechanism to accomplish the detection of atrial fibrillation using a signal-lead ECG signal. The proposed approach contains three levels of attention model including the wave, heartbeat and window levels. The attention mechanisms allow us to interpret the detection results with the high resolution. The experiment results on two different database including the the MIT-BIH Atrial fibrillation (MIT-BIH AFIB) and the PhysioNet Computing in Cardiology Challenge 2017 dataset (AFDB17) databases reveal that our method achieves the state-of-the-art performance and outperforms the the existing algorithms. Furthermore, the visualizations show that the three attention levels can consider different weights for various parts the provided signal when assigning a label to the input ECG signal. Moreover, we demonstrated that the pointed artifacts of signals by the model were clinically meaningful. A future research direction is to apply the proposed method to other ECG leads and other arrhythmias to extract new patterns that might be reasonable clinical features in the detection of an arrhythmia.
\label{sec:conclude}
\section{Acknowledgments}
This study is based upon work supported by the National Science Foundation under Grant Number 1657260. Research reported in this publication was supported by the National Institute On Minority Health And Health Disparities of the National Institutes of Health under Award Number
U54MD012388.

\bibliographystyle{ACM-Reference-Format}
\bibliography{sample-base}

\end{document}